\documentclass[12pt]{article}
\usepackage[a4paper, hmargin=2cm, vmargin=2.5cm]{geometry}
\usepackage{amsthm,amsmath,amsfonts, amssymb}
\usepackage{graphicx}        

\newcommand{\btheta}  {\boldsymbol\theta}
\newcommand{\bmu}     {\boldsymbol\mu}
\newcommand{\bdelta}  {\boldsymbol\delta}
\newcommand{\bPsi}    {\boldsymbol\Psi}
\newcommand{\bLambda} {\boldsymbol\Lambda}
\newcommand{\bSigma}  {\boldsymbol\Sigma}
\newcommand{\bDelta}  {\boldsymbol\Delta}
\newcommand{\bOmega}  {\boldsymbol\Omega}

\newcommand{\kth}     {^{(k)}}

\newcommand{\qhjk}    {\boldsymbol q_{hj}\kth}
\newcommand{\byj}     {\boldsymbol y_j}
\newcommand{\dkyj}    {d\kth(\byj)}
\newcommand{\dhkyj}   {d_h\kth(\byj)}

\newcommand{\be}	  {\mbox{\boldmath $e$}}
\newcommand{\by}	  {\mbox{\boldmath $y$}}
\newcommand{\Sahu}{Sahu et al. (2003)}

\begin{document}

\title{On the fitting of mixtures of multivariate skew $t$-distributions 
via the EM algorithm}

\author{ Sharon Lee, Geoffrey J. McLachlan	 \\ 
Department of mathematics, the University of Queensland,\\ Brisbane, Australia}

\date{}
\maketitle

\begin{abstract}
We show how the expectation-maximization (EM) algorithm can be applied exactly for 
the fitting of mixtures of a general multivariate skew $t$ (MST) distributions, 
eliminating the need for computationally expensive Monte Carlo estimation. Finite 
mixtures of MST distributions have proven to be useful in modelling heterogeneous 
data with asymmetric and heavy tail behaviour. Recently, they have been exploited 
as an effective tool for modelling flow cytometric data. However, without 
restrictions on the the characterizations of the component skew $t$-distributions, 
Monte Carlo methods have been used to fit these models. In this paper, we show how 
the EM algorithm can be implemented for the iterative computation of the maximum 
likelihood estimates of the model parameters without resorting to Monte Carlo 
methods for mixtures with unrestricted  MST components. The fast calculation of 
semi-infinite integrals on the E-step of the EM algorithm is effected by noting 
that they can be put in the form of moments of the truncated multivariate 
$t$-distribution, which subsequently can be expressed in terms of the 
non-truncated form of the $t$-distribution function for which fast 
algorithms are available. We demonstrate the usefulness of the proposed methodology by some applications to three real data sets. 
\end{abstract}

\section{Introduction}
\label{intro} 

Finite mixture distributions have become increasingly popular in the modelling 
and analysis of data due to their flexibility. This use of finite mixture 
distributions to model heterogeneous data has undergone intensive development in 
the past decades, as witnessed by the numerous applications in various scientific 
fields such as bioinformatics, cluster analysis, genetics, information processing, 
medicine, and pattern recognition. Comprehensive surveys on mixture models and 
their applications can be found, for example, in the monographs by Everitt and 
Hand (1981), Titterington, Smith, and Markov (1985), Lindsay (1995), McLachlan and 
Basford (1988), and McLachlan and Peel (2000), among others; see also the papers 
by Banfield and Raftery (1993) and Fraley and Raftery (1999).

Mixtures of multivariate $t$-distributions, as proposed by McLachlan and Peel
(1998, 2000), provide extra flexibility over normal mixtures. The thickness of 
tails can be regulated by an additional parameter -- the degrees of freedom, thus 
enabling it to accommodate outliers better than normal distributions. However, in 
many practical problems, the data often involve observations whose distributions 
are highly asymmetric as well as having longer tails than the normal, for example, 
datasets from flow cytometry (Pyne et al., 2009). Azzalini (1985) introduced the 
so-called skew-normal (SN) distribution for modelling symmetry in data sets. 
Following the development of the SN and skew $t$-mixture models by Lin, Lee, and 
Yen (2007), and Lin, Lee, and Hsieh (2007), respectively, Basso et al. (2010) 
studied a class of mixture models where the components densities are scale 
mixtures of skew-normal distributions introduced by Branco and Dey (2001), which 
include the classical skew-normal and skew $t$-distributions as special cases. 
Recently, Cabral, Lachos, and Prates (2012) have extended the work of Basso et al. 
(2010) to the multivariate case.

In a study of automated flow cytometry analysis, Pyne et al. (2009) proposed a 
finite mixture of multivariate skew $t$-distributions based on a `restricted' 
variant of the skew $t$-distribution introduced by Sahu, Dey, and Branco (2003). 
Lin (2010) considered a similar approach, but working with the original 
(unrestricted) characterization by \Sahu. However, with this more general 
formulation, maximum likelihood (ML) estimation via the EM algorithm (Dempster, 
Laird, and Rubin, 1977) can no longer be implemented in closed form due to the 
intractability of some of the conditional expectations involved on the E-step. To 
work around this, Lin (2010) proposed a Monte Carlo (MC) version of the E-step. 
One potential drawback of this approach is that the model fitting procedure relies 
on MC estimates which can be computationally expensive. 

In this paper, we show how the EM algorithm can be implemented exactly to 
calculate the ML estimates of the parameters for the (unrestricted) multivariate 
skew $t$-mixture model, based on analytically reduced expressions for the 
conditional expectations, suitable for numerical evaluation using readily 
available software. A key factor in being able to compute the integrals quickly by 
numerical means is the recognition that they can be expressed as moments of a 
truncated multivariate $t$-distribution, which in turn can be 
expressed in terms of the distribution function of a (non-truncated) multivariate 
central $t$-random vector, for which fast programs already exist. We show that the 
proposed algorithm is highly efficient compared to the version with a MC E-step. 
It produces highly accurate results for which, if MC were to achieve comparable 
accuracy, a large number of draws would be necessary. 

The remainder of the paper is organized as follows. 
In Section~\ref{sec:prelim}, 
for the sake of completeness, we include a brief description of the multivariate 
skew $t$-distribution (MST) used for defining the multivariate skew $t$-mixture 
model. We also describe the truncated $t$-distribution in the multivariate case, 
critical for the swift evaluation of the integrals on the E-step occurring in the 
calculation of some of the conditional expectations. Section \ref{sec:3} presents 
the development of an EM algorithm for obtaining ML estimates for the MST 
distribution. In the following section, the finite mixture of MST (FM-MST) 
distributions is defined. Section \ref{sec:5} presents an implementation of the EM 
algorithm to the fitting of the FM-MST model. An approximation to the observed 
information matrix is discussed in Section \ref{sec:6}. Finally, we present some 
applications of the proposed methodology in Section \ref{sec:7}.  

\section{Preliminaries}
\label{sec:prelim}

We begin by defining the multivariate skew $t$-distribution and 
briefly describing some related properties. Some alternative versions of the 
distribution are also discussed. Next, we introduce the truncated multivariate
$t$-distribution and provide some formulas for computing its moments. These 
expressions are crucial for the swift evaluation of the conditional expectations 
on the E-step to be discussed in the next section.

\subsection{The Multivariate Skew $t$-Distribution}
\label{sec:2.1}

Following \Sahu, a random vector $\boldsymbol Y$ is said to follow a 
$p$-dimensional (unrestricted) skew $t$-distribution with $p \times 1$ location vector $\bmu$, $p 
\times p$ scale matrix $\bSigma$, $p \times 1$ skewness vector $\bdelta$, and 
scalar degrees of freedom $\nu$, if its density is given by 

\begin{equation}
f_p(\boldsymbol y; \bmu, \bSigma, \bdelta, \nu) = 2^p t_{p, \nu}\left(\boldsymbol 
y; \bmu, \bOmega\right) T_{p, \nu+p}\left(\boldsymbol y^*; \boldsymbol 0, 
\bLambda\right),
\label{MST}
\end{equation}
where 
\begin{align} 
\bDelta &= \mbox{diag}(\bdelta), \notag\\
\bOmega &= \bSigma + \bDelta \bDelta^T, \notag\\
\boldsymbol y^* &= \boldsymbol q \sqrt{\frac{\nu+p}{\nu+d\left(\boldsymbol 
y\right)}}, \notag\\
\boldsymbol q &= \bDelta^T\bOmega^{-1} (\boldsymbol y-\bmu), \notag\\
d\left(\boldsymbol y\right) &= (\boldsymbol y-\bmu)^T \bOmega^{-1}(\boldsymbol 
y-\bmu), \notag\\
\bLambda &= \boldsymbol I_p - \bDelta^T \bOmega^{-1} \bDelta. \notag
\end{align}
Here the operator $\mbox{diag}(\bdelta)$ denotes a diagonal matrix with diagonal 
elements specifed by the vector $\bdelta$. Also, we let $t_{p, \nu}(.; \bmu, 
\bSigma)$ be the $p$-dimensional $t$-density with location vector $\bmu$, 
scale matrix $\bSigma$, and degrees of freedom $\nu$, and $T_{p,\nu} (.;\bmu, 
\bSigma)$ the corresponding (cumulative) distribution function. The notation 
$\boldsymbol Y \sim \mbox{ST}_{p, \nu}(\bmu, \bSigma, \bdelta)$ will be used. Note 
that when $\bdelta = \boldsymbol 0$, 
(\ref{MST}) reduces to the 
symmetric $t$-density $t_{p,\nu}(\boldsymbol y; \bmu, \bSigma)$. 
Also, when $\nu \rightarrow \infty$, we obtain the skew normal distribution. 

The MST distribution admits a convenient hierarchical form,
\begin{eqnarray}
\boldsymbol Y \mid \boldsymbol u, w &\sim& N_p\left(\bmu + {\bDelta} \boldsymbol 
u, \textstyle\frac{1}{w} {\bSigma}\right), \nonumber\\
\boldsymbol U \mid w &\sim&	HN_p\left(\boldsymbol 0, \frac{1}{w} 
\boldsymbol I_p\right), \nonumber\\
	W		&\sim&	\mbox{gamma}\left(\frac{\nu}{2}, \frac{\nu}{2}\right), \nonumber\\
\label{MST_H}
\end{eqnarray} 
where $\boldsymbol I_p$ is the $p \times p$ identity matrix, $N_k(\bmu, \bSigma)$ 
denotes the multivariate normal distribution with mean $\bmu$ and covariance 
matrix $\bSigma$, $HN_p(\boldsymbol 0,\bSigma)$ represents the $p$-dimensional 
half-normal distribution with mean $\boldsymbol 0$ and scale matrix $\bSigma$, and 
$\mbox{gamma}(\alpha, \beta)$ is the Gamma distribution with mean $\alpha/\beta$. 

We observe from (\ref{MST_H}) that the MST distribution (\ref{MST}) has the 
following stochastic representation. Suppose that conditional on the value $w$ of 
the gamma random variable $W$,
\begin{equation}
\left(\begin{array}{c}\boldsymbol U_0\\ \boldsymbol U\end{array}\right)
\sim N_p \left(\left(\begin{array}{c}\bmu \\ 
\boldsymbol 0\end{array}\right), \left(\begin{array}{cc} \bSigma/w & \boldsymbol 0 
\\ \boldsymbol 0 & \boldsymbol I_p / w 
\end{array}\right)\right),
\label{MST_C}
\end{equation} 
where $\boldsymbol I_p$ denotes the $p \times p$ identity matrix, 
$\boldsymbol 0$ denotes the zero vector of appropriate dimension, and 
$\boldsymbol U_0$ is a $p$-dimensional random vector. Then 
\begin{equation}
\boldsymbol Y = \bDelta \left|\boldsymbol U\right| + \boldsymbol 
U_0
\label{MST_S}
\end{equation} 
has the multivariate skew $t$-distribution density (\ref{MST}). In the above, 
$\left|\boldsymbol U\right|$ denotes the vector whose $i$th element is the 
magnitude of the $i$th element of the  vector $\boldsymbol U$. 
It is important to note that, 
although also known as the multivariate skew $t$-distribution, the versions 
considered by Azzalini and Dalla Valle (1996), Gupta (2003), and Lachos, Ghosh, 
and Arellano-Valle (2010), among others, are different from (\ref{MST}). These 
versions are simpler in that the skew $t$-density is defined in terms involving 
only the univariate $t$-distribution function instead of the multivariate form of 
the latter as used in (\ref{MST}). Recently, Pyne et al. (2009) proposed a 
simplified version of the skew $t$-density given by (\ref{MST}) by replacing the 
term $\bDelta \left|\boldsymbol U\right|$ in (\ref{MST_S}) by the term $\bdelta 
\left|U\right|$, where $U$ is a univariate central $t$-random variable with $\nu$ degrees of freedom, 
leading to the reduced skew $t$-density:
\begin{equation}
2^p t_{p,\nu}\left(\by; \bmu, \bSigma\right) 
T_{1,\nu+p}\left(y_1^*; 0, 1\right),
\label{MST_Sam}
\end{equation} 
where $y_1^* = \left[\left(\nu+p\right)/\left(\nu+d\left(\by\right)\right)\right]^{\frac{1}{2}} \bdelta^T \bOmega^{-1} (\by-\bmu)$. 
We shall refer to this characterization of skew $t$-distribution as the `restricted' multivariate skew $t$ (rMST)distribution. 
One immediate consequence of this type of `simplification' is that the correlation 
structure of the original symmetric model is affected by the introduction of 
skewness, whereas for (\ref{MST}) the correlation structure remains the same, as 
noted in Arellano-Valle, Bolfarine, and Lachos (2007). Nevertheless, one major 
advantage of having simplified forms like (\ref{MST_Sam}) is that calculations on 
the E-step can be expressed in closed form. However, the form of skewness is 
limited in these characterizations. Here, we extend their approach to the more 
general form of the skew $t$-density as proposed by \Sahu.

\subsection{The truncated multivariate $t$-distribution}
\label{sec:2.2} 

Let $\boldsymbol X$ be a $p$-dimensional random variable having a multivariate 
$t$-distribution with location vector $\bmu$, scale matrix $\bSigma$, and $\nu$ 
degrees of freedom. Truncating $\boldsymbol x$ to the hyperplane region 
$\mathbb{A} = \left\{\boldsymbol x \leq \boldsymbol a,\; \boldsymbol a 
\in\mathbb{R}^p\right\}$, where $\boldsymbol x \leq \boldsymbol a$ means each 
element $x_i = (\boldsymbol x)_i$ is less than or equal to $a_i=(\boldsymbol a)_i$ 
for $i = 1, \ldots, p$, results in a right-truncated $t$-distribution whose 
density is given by
\begin{equation}
f_{\mathbb{A}}(\boldsymbol x; \bmu, \bSigma, \nu) = 
T_{p,\nu}^{-1} \left(\boldsymbol a;\bmu,\bSigma\right) 
t_{p,\nu}\left(\boldsymbol x;\bmu, \bSigma, \nu\right), 
\;\;\;\; \boldsymbol x \in \mathbb{A}.
\label{TTden}
\end{equation}

For a random vector $\boldsymbol X$ with density (\ref{TTden}), we write 
$\boldsymbol X \sim tt_{p,\nu}\left(\bmu, \bSigma; \mathbb{A} \right)$. For our 
purposes, we will be concerned with the first two moments of $\boldsymbol X$, 
specifically $E(\boldsymbol X)$ and $E(\boldsymbol X \boldsymbol X^T)$. Explicit 
formulas for the truncated central $t$-distribution in the univariate case 
$tt_{1,\nu}\left(0, \sigma^2; \mathbb{A}\right)$ were provided by O'Hagan (1973), 
who expressed the moments in terms of the non-truncated $t$-distribution. The  
multivariate case was studied in O'Hagan (1976), but still considering the central 
case only. We will generalize these results to the case with non-zero location vector here.

Before presenting the expressions, it will be convenient to introduce some 
notation. Let $\boldsymbol x$ be a vector, where $x_i$ denotes the $i$th element 
and $\boldsymbol x_{ij}$ is a two-dimensional vector with elements $x_i$ and 
$x_j$. Also, $\boldsymbol x_{-i}$ and $\boldsymbol x_{-ij}$ represents the $(p-1)$ 
and $(p-2)$-dimensional vector, respectively, with the corresponding elements 
removed. For a matrix $\boldsymbol X$, $x_{ij}$ denotes the $ij$th element, and 
$\boldsymbol X_{ij}$ defines the $2\times 2$ matrix consisting of the elements 
$x_{ii}$, $x_{ij}$, $x_{ji}$ and $x_{jj}$. $\boldsymbol X_{-i}$ is created by 
removing the $i$th row and column from $\boldsymbol X$. Similarly, $\boldsymbol 
X_{-ij}$ is the $(p-2)$-dimensional square matrix resulting from the removal of the $i$th and 
$j$th row and column from $\boldsymbol X$. Lastly, $\boldsymbol X_{(ij)}$ is the 
$i$th and $j$th column of $\boldsymbol X$ with the elements of $\boldsymbol 
X_{ij}$ removed, yielding a $(p-2)\times 2$ matrix. We now proceed to the 
expressions for the first two moments of $\boldsymbol X$. \newline

With some effort, one can show that the first moment of (\ref{TTden}) is
\begin{align}
E\left(\boldsymbol X\right) &= \bmu - c^{-1}\bSigma \boldsymbol \xi 
	= \bmu - \bmu^*, 
\label{EX}
\end{align}
where $c = T_{p,\nu}\left(\boldsymbol a-\bmu;\boldsymbol 0, \bSigma\right)$, and 
$\boldsymbol \xi$ is a $p \times 1$ vector with elements
\begin{align}
\xi_i &= \left(2\pi \sigma_{ii}\right)^{-\frac{1}{2}} 
	\left(\frac{\nu}{\nu+\sigma_{ii}^{-1} 
	(a_i-\mu_i)^2}\right)^{(\frac{\nu-1}{2})} 
	\frac{\Gamma\left(\frac{\nu-1}{2}\right)}{\Gamma\left(\frac{\nu}{2}\right)} 
	\sqrt{\frac{\nu}{2}} T_{p-1,\nu-1}\left(a^*;\boldsymbol 0,\bSigma^*\right), \notag
\end{align}
for $i = 1, \ldots, p$, and where 
\begin{align}
\boldsymbol a^* &= \left(\boldsymbol a_{-i}-\bmu_{-i}\right) - \left(\boldsymbol 
a_{i}-\bmu_{i}\right)\sigma_{ii}^{-1} 
\bSigma_{(i)} \notag
\intertext{and}
\bSigma^* &= \left(\frac{\nu+\sigma_{ii}^{-1} 
\left(a_i-\mu_i\right)^2}{\nu-1}\right) \left(\bSigma_{-i} - 
\frac{1}{\sigma_{ii}}\bSigma_{(i)}\bSigma_{(i)}^T\right). \notag
\end{align} \newline

The second moment is given by
\begin{eqnarray}
E\left(\boldsymbol X\boldsymbol X^T\right) &=& \bmu \bmu^T - \bmu \bmu^{*^T} - 
\bmu^*\bmu^T - c^{-1} \bSigma \boldsymbol H \bSigma \nonumber\\
	&	&	+ c^{-1} \left(\textstyle\frac{\nu}{\nu-2}\right) 
	T_{p,\nu-2}\left(\boldsymbol 	a-\bmu;\boldsymbol 0, 
	\left(\textstyle\frac{\nu}{\nu-2}\right)\bSigma\right) \bSigma, 
	\label{EXX}
\end{eqnarray} 
where $\boldsymbol H$ is a $p \times p$ matrix with off-diagonal elements   
\begin{equation}
h_{ij} = - \frac{1}{2\pi\sqrt{\sigma_{ii}\sigma_{jj}-\sigma_{ij}^2}} 
\left(\frac{\nu}{\nu-2}\right)\left(\frac{\nu}{\nu^*}\right)^{\frac{\nu}{2}
-1} T_{p-2, \nu-2}\left(\boldsymbol a^{**}; \boldsymbol 0,\bSigma^{**}\right), 
\;\; i \neq j, \nonumber 
\end{equation}  
and diagonal elements, 
\begin{align}
h_{ii} &= \sigma_{ii}^{-1}(a_i-\mu_i) \xi_i - \sigma_{ii}^{-1} \sum_{j\neq i} 
\sigma_{ij} h_{ij}, \notag\\
\intertext{and} 
\nu^* &= \nu + \left(\boldsymbol a_{ij}-\bmu_{ij}\right)^T \bSigma_{ij}^{-1} 
\left(\boldsymbol a_{ij}-\bmu_{ij}\right), \notag\\
\boldsymbol a^{**} &= \left(\boldsymbol a_{-ij}-\bmu_{-ij}\right) - \bSigma_{(ij)} 
\bSigma_{ij}^{-1} \left(\boldsymbol a_{ij}-\bmu_{ij}\right), \notag\\
\bSigma^{**} &= \frac{\nu^*}{\nu-2} \left(\bSigma_{-ij} - 
\bSigma_{(ij)} \bSigma_{ij}^{-1} \bSigma_{(ij)}^T\right). 
\notag
\end{align}

It is worth noting that evaluation of the expressions (\ref{EX}) and (\ref{EXX}) 
rely on algorithms for computing the multivariate central $t$-distribution 
function for which highly efficient procedures are readily available in many 
statistical packages. For example, an implementation of Genz's algorithm 
(Genz and Bretz, 2002; Kotz and Nadarajah, 2004) is provided by the 
\texttt{mvtnorm} package available from the R website.

\section{ML Estimation for the MST Distribution}
\label{sec:3}

In this section, we describe an EM algorithm for the ML estimation of the MST 
distribution specified by (\ref{MST}). To apply the EM algorithm, the observed 
data vector $\boldsymbol y = \left(\boldsymbol y_1^T, \ldots, \boldsymbol 
y_n^T\right)^T$ is regarded as incomplete, and we introduce two latent variables 
denoted by $\boldsymbol u$ and $w$, as defined by (\ref{MST_H}). We let $\btheta$ 
be the parameter containing the elements of the location parameter $\bmu$, the 
distinct elements of the scale matrix $\bSigma$, the elements of the skew 
parameter $\bdelta$, and the degrees of freedom $\nu$. It follows that the 
complete-data log-likelihood function for $\btheta$ is given by 
\begin{eqnarray}
\log L_c(\btheta; \boldsymbol y, \boldsymbol u, w) &=& K - 
\textstyle\frac{1}{2} n
\log\left|\bSigma\right| - n\log\Gamma 
\left(\textstyle\frac{1}{2}\nu\right) + \textstyle\frac{1}{2}n\nu 
\log\left(\textstyle\frac{1}{2}\nu\right) \nonumber\\
	&	&	- \textstyle\frac{1}{2}w\left(d\left(\boldsymbol y\right) + 
	\left(\boldsymbol u-\boldsymbol q\right)^T \bLambda^{-1} 
	\left(\boldsymbol u-\boldsymbol q\right)\right) \nonumber\\
	&	& + \left(\textstyle\frac{1}{2}\nu+p-1\right)\log(w),
\label{Lc}
\end{eqnarray}
where $K$ does not depend on $\btheta$. 

The implementation of the EM algorithm requires alternating repeatedly the E- and 
M-steps until convergence in the case where the sequence of the log likelihood 
values ${L(\btheta^{(k)})}$ is bounded above. Here $\btheta^{(k)}$ denotes the 
value of $\btheta$ after the $k$th iteration.

On the $(k+1)$th iteration, the E-step requires the calculation of the conditional 
expectation of the complete-data log likelihood given the observed data 
$\boldsymbol y$, using the current estimate $\btheta^{(k)}$ for $\btheta$. That 
is, we have to calculate the so-called $Q$-function defined by
\begin{equation}
	Q(\btheta; \btheta^{(k)}) 
	= E_{\btheta^{(k)}} \left\{\log L_c(\btheta; \boldsymbol y, 
	\boldsymbol u, w) \mid \boldsymbol y \right\},
\label{Qfun}
\end{equation}
where $E_{\btheta^{(k)}}$ denotes the expectation operator, using $\btheta^{(k)}$ 
for $\btheta$. This, in effect, requires the calculation of the conditional 
expectations 
\begin{align}
e_{1,j}^{(k)} &= E_{\btheta^{(k)}} \left\{\log(W_j)\mid \byj\right\}, 
\notag\\ 
e_{2,j}^{(k)} &= E_{\btheta^{(k)}} \left\{W_j\mid\boldsymbol y_j \right\}, 
\notag\\ 
\be_{3,j}\kth &= E_{\btheta^{(k)}} \left\{W_j\boldsymbol U_j\mid\byj\right\}, 
\notag\\ 
\be_{4,j}^{(k)}&=E_{\btheta^{(k)}} \left\{W_j\boldsymbol U_j\boldsymbol U_j^T \mid 
\byj\right\}. 
\notag 
\end{align}

Note that the $Q$-function does not admit a closed form expression for this 
problem, due to the conditional expectations $e_{1,j}^{(k)}$, $\be_{3,j}^{(k)}$, 
and $\be_{4,j}^{(k)}$ not being able to be evaluated in 
closed form. \newline

Concerning the calculation of the expectation $e_{1,j}^{(k)}$, the conditional 
density of $W_j$ given $\byj$, is given by
\begin{equation}
f(w_j\mid\byj) = \frac{\Gamma\left(w_j; \frac{\nu\kth+p}{2}, 
\frac{\nu\kth+\dkyj}{2}\right) \Phi_p\left(\boldsymbol q_j^{(k)} 
\sqrt{w_j}; \boldsymbol 0, \bLambda^{(k)}\right)} 
{T_{p,\nu\kth+p}\left(\byj^{*(k)}; \boldsymbol 0, \bLambda^{(k)}\right)},
\label{W_Y}
\end{equation}
where 
\begin{eqnarray}
\byj^{*(k)} &=& \boldsymbol q_j^{(k)} \sqrt{\frac{\nu^{(k)}+p} 
{\nu^{(k)}+\dkyj}}, \nonumber\\
\boldsymbol q_j^{(k)} &=& \bDelta^{(k)^T} \bOmega^{(k)^{-1}} 
\left(\byj-\bmu^{(k)}\right), \nonumber\\
\dkyj &=& \left(\byj-\bmu^{(k)}\right)^T \bOmega^{(k)^{-1}} \left(\byj - 
\bmu^{(k)}\right), \nonumber
\end{eqnarray}
and $\boldsymbol 0$ is the zero vector of appropriate dimension. 

The conditional expectation $E_{\btheta^{(k)}} \left\{\log(W_j) \mid\byj\right\}$ 
can be reduced to 
\begin{eqnarray}
e_{1,j}^{(k)} &=&	\left(\frac{\nu^{(k)}+p}{\nu^{(k)}+\dkyj}\right) 
\frac{T_{p,\nu^{(k)}+p+2}\left(\boldsymbol q_j^{(k)} 
\sqrt{\frac{\nu^{(k)}+p+2}{\nu^{(k)}+\dkyj}}; 
\boldsymbol 0, \bLambda^{(k)}\right)} {T_{p,\nu^{(k)}+p}\left(\byj^{*(k)}; 
\boldsymbol 0, \bLambda^{(k)}\right)} 
\nonumber\\
	&	&	- \log\left(\frac{\nu^{(k)}+\dkyj}{2}\right) - 
	\left(\frac{\nu^{(k)}+p}{\nu^{(k)}+\dkyj}\right) + 
	\psi\left(\frac{\nu^{(k)}+p}{2}\right) + S,
\label{E1a}
\end{eqnarray}
where the last term $S$ is given by 
\begin{eqnarray}
S &=&	\psi\left(\frac{\nu^{(k)}}{2}+p\right) - 
\psi\left(\frac{\nu^{(k)}+p}{2}\right) + 
\left(\frac{\nu^{(k)}+p}{\nu^{(k)}+\dkyj}\right) \nonumber\\
\nonumber\\
	&	&	- \left(\frac{\nu^{(k)}+p}{\nu^{(k)}+\dkyj}\right) 
	\frac{T_{p,\nu^{(k)}+p+2}\left(\boldsymbol q_j^{(k)} 
	\sqrt{\frac{\nu^{(k)}+p+2}{\nu^{(k)}+\dkyj}}; \boldsymbol 0, 
	\bLambda^{(k)} \right)} {T_{p,\nu^{(k)}+p}\left(\boldsymbol 
	q_j^{(k)}\sqrt{\frac{\nu^{(k)}+p}{\nu^{(k)}+\dkyj}};
\boldsymbol 0, \bLambda^{(k)}\right)}
 \nonumber\\ \nonumber\\
 	&	&	- \frac{\left[\pi\left(\nu^{(k)}+p\right)\right]^{-\frac{p}{2}} 
 	\left|\bLambda\right|^{-\frac{1}{2}}}{T_{p,\nu^{(k)}+p} 
 	\left(\boldsymbol 	q_j^{(k)} 
\sqrt{\frac{\nu^{(k)}+p}{\nu^{(k)}+\dkyj}};\boldsymbol 0, \bLambda^{(k)}\right)}	
\frac{\Gamma\left(\frac{\nu^{(k)}}{2}+p\right)}{\Gamma\left(\frac{\nu^{(k)}+p}{2}
\right)} S_{1,j}^{(k)}, 
 	\label{S}	
\end{eqnarray}
and $S_{1,j}^{(k)}$ is an integral given by
\begin{eqnarray}
S_{1,j}^{(k)} &=& \int_{-\infty}^{\left[\boldsymbol q_{j}^{(k)}\right]_1} 
\int_{-\infty}^{\left[\boldsymbol q_{j}^{(k)}\right]_2} \ldots 
\int_{-\infty}^{\left[\boldsymbol q_{j}^{(k)}\right]_p} 
log\left(1+\frac{\boldsymbol s^T{\bLambda}^{{(k)}^{-1}}\boldsymbol 
s}{\nu^{(k)}+\dkyj}\right) \nonumber\\
	&	& \left[1+\frac{\boldsymbol s^T{\bLambda}^{{(k)}^{-1}}
\boldsymbol s}{\nu^{(k)}+d^{(k)}(\byj)}\right] 
	^{-\left(\frac{\nu^{(k)}}{2}+p\right)} ds_1 
	ds_2 \ldots ds_p,
	\label{S1}
\end{eqnarray}
and $\psi(\cdot)$ denotes the Digamma function. 

Combining (\ref{E1a}) and (\ref{S}), $e_{1,j}^{(k)}$ can be reduced to
\begin{eqnarray}
e_{1,j}^{(k)} &=& \psi\left(\frac{\nu^{(k)}}{2}+p\right) - 
\log\left(\frac{\nu^{(k)}+\dkyj}{2}\right) \nonumber\\
	&	&	- T_{p,\nu^{(k)}+p}^{-1}\left(\boldsymbol q_j^{(k)} 
	\sqrt{\frac{\nu^{(k)}+p}{\nu^{(k)}+\dkyj}}; 
\boldsymbol 0, \bLambda^{(k)}\right) S_{1,j}^{(k)}.
\label{E1}
\end{eqnarray}

We note that the term $S$ will be very small in practice since it would be zero if 
we adopted a one-step late (OSL) EM algorithm (Green, 1990). In which case, there 
would be no need to calculate the multiple integral $S_{1,j}^{(k)}$ in (\ref{S}). 
Hence then, $e_{1,j}^{(k)}$ can be reduced to
\begin{eqnarray}
e_{1,j}^{(k)} &=& 
\left(\frac{\nu^{(k)}+p}{\nu^{(k)}+\dkyj}\right) 
\frac{T_{p,\nu^{(k)}+p+2}\left(\boldsymbol 
q_j^{(k)} \sqrt{\frac{\nu^{(k)}+p+2}{\nu^{(k)}+\dkyj}}; \boldsymbol 0, 
\bLambda^{(k)}\right)} 
{T_{p,\nu^{(k)}+p}\left(\byj^{*(k)}; \boldsymbol 0, \boldsymbol 
\Lambda^{(k)}\right)} \nonumber\\
	&	&	- \log\left(\frac{\nu^{(k)}+\dkyj}{2}\right) - 
	\left(\frac{\nu^{(k)}+p}{\nu^{(k)}+\dkyj}\right) + 
	\psi\left(\frac{\nu^{(k)}+p}{2}\right).
\label{E1apprx}
\end{eqnarray}

It can be easily shown that $e_{2,j}^{(k)}$ can be written in closed form 
(see, for example, Lin (2010)), given by
\begin{equation}
e_{2,j}^{(k)} = \left(\frac{\nu^{(k)}+p}{\nu^{(k)}+\dkyj}\right) 
\frac{T_{p,\nu^{(k)}+p+2}\left(\boldsymbol q_j^{(k)} 
\sqrt{\frac{\nu^{(k)}+p+2}{\nu^{(k)}+\dkyj}}; 
\boldsymbol 0, \bLambda^{(k)}\right)}{T_{p,\nu^{(k)}+p}\left(\boldsymbol 
y_j^{*(k)}; \boldsymbol 0, \bLambda^{(k)}\right)}.
\label{E2}
\end{equation}

To obtain $\be_{3,j}^{(k)}$ and $\be_{4,j}^{(k)}$, first note that the joint 
density of $\byj$, $\boldsymbol u_j$, and $w_j$ is given by 
\begin{eqnarray}
f(\byj, \boldsymbol u_j, w_j) &=& \pi^{-p} 
\Gamma\left(\frac{\nu^{(k)}}{2}\right)^{-1}
\left(\frac{\nu^{(k)}}{2}\right)^{\left(\frac{\nu^{(k)}}{2}\right)} 
w_j^{\left(\frac{\nu^{(k)}}{2}+p-1\right)} \nonumber\\
	&	& e^{-\frac{w_j}{2} 
\left[\nu^{(k)} + \dkyj + \left(\boldsymbol u_j-\boldsymbol q_j^{(k)}\right)^T 
\bLambda^{(k)^{-1}}\left(\boldsymbol u_j-\boldsymbol q_j^{(k)}\right)\right]}. 
\label{YUW}
\end{eqnarray} 

Using Bayes' rule, the conditional density of $\boldsymbol u_j$ and $w_j$ 
given $\byj$ can be written as
\begin{equation}
f(\boldsymbol u_j, w_j \mid \byj) = \frac{w_j^{\frac{p}{2}} 
\Gamma\left(w_j;\frac{\nu^{(k)}+p}{2},\frac{d^{(k)}\left(\boldsymbol 
y_j\right)}{2}\right) e^{-\frac{w_j}{2}\left(\boldsymbol 
u_j-\boldsymbol q_j^{(k)}\right)^T\bLambda^{(k)^{-1}}\left(\boldsymbol 
u_j-\boldsymbol q_j^{(k)}\right)}} {(2\pi)^{\frac{p}{2}} 
\left|\bLambda^{(k)}\right|^{\frac{1}{2}} T_{p,\nu^{(k)}u+p}\left(\boldsymbol 
q_j^{(k)}\sqrt{\frac{\nu^{(k)}+p}{\nu^{(k)}+d^{(k)}\left(\boldsymbol 
y_j\right)}}; \boldsymbol 0, \bLambda^{(k)}\right)}. \nonumber
\label{UW_Y}
\end{equation} 

From (\ref{UW_Y}), standard conditional expectation calculations yield 
\begin{eqnarray}
\be_{3,j}^{(k)} = 
\left(\frac{\nu^{(k)}+p}{\nu^{(k)}+d^{(k)}\left(\boldsymbol 
y_j\right)}\right) \frac{T_{\nu^{(k)}+p+2}\left(\boldsymbol q_j^{(k)}; 
\boldsymbol 0, \left(\frac{\nu^{(k)}+d^{(k)}(\byj)} 
{\nu^{(k)}+p+2}\right)\bLambda^{(k)}\right)} 
{T_{p,\nu^{(k)}+p}\left(\boldsymbol 
y_j^{*(k)}; \boldsymbol 0, \bLambda^{(k)}\right)} \boldsymbol 
S_{2,j}^{(k)}
	= e_{2,j}^{(k)} \boldsymbol S_{2,j}^{(k)}, \nonumber\\
\label{E3}
\end{eqnarray}
where $\boldsymbol S_{2,j}^{(k)}$ represents the expected value of a truncated 
$p$-dimensional $t$-variate $\boldsymbol X_j$, which is distributed as,
\begin{equation}
\boldsymbol X_j \sim tt_{p,\nu^{(k)}+p+2}\left(\boldsymbol 
q_j^{(k)}, \left(\frac{\nu^{(k)}+d^{(k)}(\byj)} {\nu^{(k)}+p+2}\right) \bLambda^{(k)}; 
\mathbb{R}^+\right).
\label{X}
\end{equation} 
That is, the random vector $\boldsymbol X_j$ is truncated to lie in the positive 
hyperplane $\mathbb{R}^+$.

Analogously, $\be_{4,j}^{(k)}$ can be reduced to
\begin{eqnarray}
\be_{4,j}^{(k)} = \left(\frac{\nu^{(k)}+p}{\nu^{(k)}+d^{(k)}(\boldsymbol 
y_j)}\right) \frac{T_{p, \nu^{(k)}+p+2}\left(\boldsymbol q_j^{(k)}; 
\boldsymbol 0, \left(\frac{\nu^{(k)}+\dkyj} 
{\nu^{(k)}+p+2}\right)\bLambda^{(k)}\right)} {T_{p,\nu^{(k)}+p}\left(\byj^{*(k)}; 
\boldsymbol 0, \bLambda^{(k)}\right)} \boldsymbol S_{3,j}^{(k)}
	= e_{2,j}^{(k)} \boldsymbol S_{3,j}^{(k)}, \nonumber\\
\label{E4}
\end{eqnarray}   
where $\boldsymbol S_{3,j}^{(k)}$ represents the second moment of 
$\boldsymbol X_j$. The truncated moments $\boldsymbol S_{2,ij}^{(k)}$ and 
$\boldsymbol S_{3,ij}^{(k)}$ can be swiftly evaluated using the expressions 
(\ref{EX}) and (\ref{EXX}) in Section \ref{sec:2.2}.    

\subsection{M-step}
\label{sec:3.2} 

On the $(k+1)$th iteration, the M-step consists of the maximization of the 
$Q$-function (\ref{Qfun}) with respect to $\btheta$. For easier computation, we 
employ the ECM extension of the EM algorithm, where the M-step is replaced by four 
conditional--maximization (CM)-steps, corresponding to the decomposition of 
$\btheta$ into four subvectors, $\btheta = (\btheta_1^T, \btheta_2^T, \btheta_3^T, 
\theta_4)^T$, where $\btheta_1=\bmu$, $\btheta_2=\bdelta$, $\btheta_3$ is the 
vector containing the distinct elements of $\bSigma$, and $\theta_4 = \nu$. To 
compute $\bmu^{(k+1)}$, we maximize $Q(\bmu, \btheta_2\kth, \btheta_3\kth, 
\theta_4\kth;\btheta\kth)$ with respect to $\bmu$, and to compute 
$\bdelta^{(k+1)}$, we first update $\bmu$ to $\bmu^{(k+1)}$ and then maximize 
$Q(\bmu^{(k+1)},\bdelta,\btheta_3\kth,\theta_4\kth; \btheta\kth)$ with respect to 
$\bdelta$, and so on.

We let $\mbox{DIAG}(\boldsymbol A)$ denote the operator that 
produces a vector by extracting the diagonal elements of $\boldsymbol A$. 
Straightforward algebraic manipulations lead to the following closed form expressions for 
$\bmu^{(k+1)}$, $\bSigma^{(k+1)}$, and $\bdelta^{(k+1)}$,
\begin{align}
\bmu^{(k+1)}	&=	\frac{\sum_{j=1}^n \left[e_{2,j}^{(k)}\byj -\bDelta\kth 
\be_{3,j}^{(k)}\right]} {\sum_{j=1}^n e_{2,j}^{(k)}}, \label{MU} 
\displaybreak[0]\\
\bdelta^{(k+1)} &= \left(\bSigma^{(k)^{-1}} \odot 
	\sum_{j=1}^n \be_{4.j}^{(k)}\right)^{-1}
	\mbox{DIAG}\left(\bSigma^{(k)^{-1}} 
	\sum_{j=1}^n (\by_j-\bmu^{(k)}) \be_{3,j}^{(k)^T} \right), 
	\label{DELTA} 
\displaybreak[0]\\
\intertext{and} 
\bSigma^{(k+1)} &=	\frac{1} {n} \sum_{j=1}^n 
\left[\bDelta^{(k+1)}\be_{4,j}^{(k)^T}\bDelta^{(k+1)^T} \right.
	 - \left(\byj-\bmu^{(k+1)}\right)\be_{3,j}^{(k)^T}\bDelta^{(k+1)} 
	\notag\\
	& + \left(\byj-\bmu^{(k+1)}\right) \left(\byj-\bmu^{(k+1)}\right)^T 
	e_{2,j}^{(k)} \left. - {\bDelta}^{(k+1)}\be_{3,j}^{(k)} 
	\left(\byj-\bmu^{(k+1)}\right)^T \right], 
\label{SIGMA}
\end{align}
where $\odot$ denotes the Hadamard or element-wise product, and $\bDelta^{(k+1)} = \mbox{diag} \left(\bdelta^{(k+1)}\right)$. 

An updated estimate of the degrees of freedom $\nu^{(k+1)}$ is obtained by solving 
the equation
\begin{equation}
\log\left(\frac{\nu^{(k+1)}}{2}\right) - 
\psi\left(\frac{\nu^{(k+1)}}{2}\right) + 1 = \frac{1}{n} \sum_{j=1}^n 
\left(e_{2,j}^{(k)}-e_{1,j}^{(k)}\right).
\label{NU}
\end{equation}
\newline
 
In summary, the ECM algorithm proceeds as follows on the $(k+1)$th
iteration: \newline\newline
{\textbf E-step:} Given $\btheta = \btheta^{(k)}$, compute the four conditional 
expectations $e_{1,j}^{(k)}$, $e_{2,j}^{(k)}$, $\be_{3,j}^{(k)}$ and 
$\be_{4,j}^{(k)}$ by using (\ref{E1}), (\ref{E2}), (\ref{E3}), and 
(\ref{E4}), respectively, for $j=1, \ldots, n$. \newline\newline
{\textbf M-step:} Update $\bmu^{(k+1)}$, $\bdelta^{(k+1)}$ , $\bSigma^{(k+1)}$ and 
by using (\ref{MU}), (\ref{DELTA}), and (\ref{SIGMA}). Calculate $\nu^{(k+1)}$ by 
solving (\ref{NU}).

\section{The Multivariate Skew $t$-Mixture Model}
\label{sec:4}

The probability density function (pdf) of a finite mixture of $g$ multivariate 
skew $t$-components, using the notation above, is given by
\begin{equation}
	f\left(\boldsymbol y; \bPsi\right) = \sum_{h=1}^g \pi_h f_p \left(\boldsymbol 
	y; \bmu_h, \bSigma_h, \bdelta_h , \nu_h \right),
\label{MSTmix} 
\end{equation}
where $f_p \left(\boldsymbol y; \bmu_h, \bSigma_h, \bdelta_h, \nu_h \right)$ 
denotes the $i$th MST component of the mixture model as defined by (\ref{MST}), 
with location parameter $\bmu_h$, scale matrix $\bSigma_h$, skew parameter 
$\bdelta_h$, and degrees of freedom $\nu_h$. The mixing proportions $\pi_h$ 
satisfy $\pi_h \geq 0$ $(h=1,\ldots,g)$ and $\sum_{h=1}^g \pi_h = 1$. We shall 
denote the model defined by (\ref{MSTmix}) by FM-MST (finite mixture of MST) 
distributions. Let $\bPsi$ contain all the unknown parameters of the FM-MST model; 
that is, $\bPsi = \left(\pi_1, \ldots, \pi_{g-1}, \btheta_1^T, \ldots, 
\btheta_g^T\right)^T$ where now $\btheta_h$ consists of the unknown parameters of 
the $i$th component density function. 
\newline

To formulate the estimation of the unknown parameters in the FM-MST model as an 
incomplete-data problem in the EM framework, a set of latent component labels 
$\boldsymbol z_j = \left(z_{1j}, \ldots, z_{gj}\right)^T$ $(j=1, \ldots, n)$ is 
introduced, where each element $z_{hj}$ is a binary variable defined as
\begin{equation}
z_{hj} = \left\{ \begin{array}{cc}1, & \mbox{if}\;\boldsymbol 
y_j,\;\mbox{belongs to component}\;i, \\ 0, & \mbox{otherwise}, \end{array} 
\right.
\label{Z}
\end{equation}  
and $\sum_{h=1}^g z_{hj} =1$ $(j=1,\ldots,n)$. Hence, the random vector 
$\boldsymbol Z_j$ corresponding to $\boldsymbol z_j$ follows a multinomial 
distribution with one trial and cell probabilities $\pi_1, \ldots, \pi_g$; that 
is, $\boldsymbol Z_j \sim \mbox{Mult}_g (1; \pi_1, \ldots,\pi_g)$. It follows that 
the FM-MST model can be represented in the hierarchical form given by
\begin{eqnarray}
\boldsymbol Y_j \mid \boldsymbol u_j, w_j, z_{hj}=1 &\sim&	
N_p\left(\bmu_h + \bDelta_h \boldsymbol u_j, \frac{1}{w_j} \bSigma_h\right), 
\nonumber\\
\boldsymbol U_j \mid w_j, z_{hj}=1 &\sim&	HN_p\left(\boldsymbol 0, \frac{1}{w_j} 
\boldsymbol I_p\right), \nonumber\\
	W_j	\mid z_{hj}=1	&\sim&	\mbox{gamma}\left(\frac{\nu_h}{2}, 
	\frac{\nu_h}{2}\right), \nonumber\\
	\boldsymbol Z_j	&\sim& \mbox{Mult}_g\left(1, \boldsymbol \pi\right),
\label{MSTMIX_H}
\end{eqnarray} 
where $\bDelta_h = \mbox{diag}\left(\bdelta_h\right)$ and $\boldsymbol 
\pi = \left(\pi_1, \ldots, \pi_g\right)^T$.

\section{ML Estimation for FM-MST Distributions}
\label{sec:5}

From the hierarchical characterization (\ref{MSTMIX_H}) of the FM-MST 
distributions, the complete-data log-likelihood function is given by
\begin{equation}
\log L_c\left(\bPsi\right) = \log L_{1c}\left(\bPsi\right) + 
\log L_{2c}\left(\bPsi\right) + \log L_{3c}\left(\bPsi\right),
\label{logL}
\end{equation}
where
\begin{align}
\log L_{1c}\left(\bPsi\right) &=	\sum_{h=1}^g \sum_{j=1}^n z_{hj} 
\log\left(\pi_h\right), \displaybreak[0]\notag\\
\log L_{2c}\left(\bPsi\right) &=	\sum_{h=1}^g \sum_{j=1}^n z_{hj} 
\left[\left(\frac{\nu_h}{2}\right) \log\left(\frac{\nu_h}{2}\right) + 
\left(\frac{\nu_h}{2}+p-1\right)\log\left(w_j\right) \right.\notag\\
	&		\left .- \log\Gamma\left(\frac{\nu_h}{2}\right) - 
	\left(\frac{w_j}{2}\right)\nu_h\right], \notag\displaybreak[0]\\
\log L_{3c}\left(\bPsi\right) &=	\sum_{h=1}^g \sum_{j=1}^n z_{hj} \left\{- 
p\log\left(2\pi\right) -  \frac{1}{2}\log\left|{\bSigma}_h\right| \right. \notag\\
	&		- \left. \frac{w_j}{2}\left[d_h\left(\byj\right) + 
	\left(\boldsymbol u_j-\boldsymbol q_{hj}\right)^T \bLambda_h^{-1} 
	\left(\boldsymbol u_j-\boldsymbol q_{hj}\right)\right]\right\}, 
\label{Lc2}
\end{align}
and where 
\begin{align}
d_h\left(\byj\right) &= \left(\byj-\bmu_h \right)^T {\bOmega}_h^{-1} 
\left(\byj-\bmu_h\right), \notag\\
\boldsymbol q_{hj} &= \bDelta_h^T {\bOmega}_h^{-1} 
\left(\byj-\bmu_h\right), \notag\\ 
{\bLambda}_h &= \boldsymbol I_p - \bDelta_h^T{\bOmega}_h^{-1}\bDelta_h, \notag\\
{\bOmega}_h &= {\bSigma}_h + \bDelta_h \bDelta_h^T. \notag
\end{align}

It is clear from (\ref{logL}) that maximization of the $Q$-function of the 
complete-data log likelihood (McLachlan and Krishnan, 2008), 
\begin{eqnarray}
Q(\bPsi; \bPsi^{(k)}) &=&	E_{\bPsi^{(k)}}\left\{\log L_c\left(\bPsi\right) \mid 
\boldsymbol y\right\}, \nonumber
\end{eqnarray} 
only requires maximization of the components functions $L_{hc}(\bPsi)$ separately 
$(h = 1, 2, 3)$. The necessary conditional expectations involved in computing the 
$Q$-function with respect to (\ref{Lc2}) are, namely, 
\begin{align}
\tau_{hj}^{(k)} &= E_{\bPsi^{(k)}}\{Z_{hj} \mid \byj\}, \notag\\
e_{1,hj}^{(k)} &= E_{\bPsi^{(k)}}\{\log(W_j) \mid \byj, z_{hj}=1\}, \notag\\ 
e_{2,hj}^{(k)} &= E_{\bPsi^{(k)}}\{W_j \boldsymbol U_j \mid \byj, z_{hj}=1\}, 
\notag\\ 
\be_{3,hj}^{(k)} &= E_{\bPsi^{(k)}} \{W_j \boldsymbol U_j \mid 
\boldsymbol y_j, z_{hj}=1\}, \notag\\
\be_{4,hj}^{(k)} &= E_{\bPsi^{(k)}} \{W_j \boldsymbol U_j \boldsymbol 
U_j^T \mid \byj, z_{hj}=1\}. 
\end{align}

The posterior probability of membership of the $h$th component by $\boldsymbol 
y_j$, using the current estimate $\bPsi^{(k)}$ for $\bPsi$, is given using Bayes' 
Theorem by
\begin{equation}
\tau_{hj}^{(k)} = \frac{\pi_h^{(k)} f_p \left(\byj; 
\bmu_h^{(k)}, {\bSigma}_h^{(k)}, \bdelta_h^{(k)}, \nu_h\kth\right)} {\sum_{h=1}^g 
\pi_h^{(k)} f_p \left(\byj;\bmu_h^{(k)}, {\bSigma}_h^{(k)}, \bdelta_h^{(k)}, 
\nu_h\kth\right)}.
\label{TAU}
\end{equation} 

The other four expectations have analogous expressions to 
their one-component counterpart given in Section \ref{sec:3}. 
They are given by 
\begin{align}
e_{1,hj}^{(k)}
	 &= \psi\left(\frac{\nu_h^{(k)}}{2}+p\right) - 
	 \log\left(\frac{\nu_h^{(k)}+d_h^{(k)}(\byj)}{2}\right) 
	 \label{e1} \\
	 &	- T_{p, \nu_h^{(k)}+p}^{-1} \left(\qhjk 
	 \sqrt{\textstyle\frac{\nu_h^{(k)}+p}{\nu_h^{(k)}+\dhkyj}}; 
	 \boldsymbol 0, \bLambda_h^{(k)}\right)  S_{1,hj}^{(k)},  
	 \notag\\
e_{2,hj}^{(k)} 
	&= \left(\frac{\nu_h^{(k)}+p}{\nu_h^{(k)}+d_h^{(k)}\left(\byj\right)}\right) 		
	\frac{T_{p,\nu_h^{(k)}+p+2}\left(\qhjk 
	\sqrt{\frac{\nu_h^{(k)}+p+2}{\nu_h^{(k)}+\dhkyj}};\boldsymbol 0, 
	{\bLambda}_h^{(k)}\right)} 
{T_{p,\nu_h^{(k)}+p}\left(y_{hj}^{*(k)}; \boldsymbol 0,
{\bLambda}_h^{(k)}\right)}, \label{e2} \displaybreak[0]\\
\be_{3,hj}^{(k)} 	&= 
\left(\frac{\nu_h^{(k)}+p}{\nu_h^{(k)}+d_h^{(k)}(\byj)}\right) T_{p, 
\nu_h^{(k)}+p}^{-1}\left(y_{hj}^{*(k)}; 
\boldsymbol 0,
\bLambda_h^{(k)}\right) \boldsymbol S_{2,ij}^{(k)}, \label{e3} 
\displaybreak[0]\\ 
\be_{4,hj}^{(k)} 	&=
\left(\frac{\nu_h^{(k)}+p}{\nu_h^{(k)}+d_h^{(k)}(\byj)}\right) T_{p, 
\nu_h^{(k)}+p}^{-1}\left(y_{hj}^{*(k)}; 
\boldsymbol 0, \bLambda_h^{(k)}\right) 
\boldsymbol S_{3,ij}^{(k)}, \label{e4} \displaybreak[0]
\end{align} 
where $S_{1,hj}^{(k)}$ is a scalar defined by
\begin{eqnarray}
S_{1,ij}^{(k)} &=& \int_{-\infty}^{\left[\qhjk\right]_1} 
\int_{-\infty}^{\left[\qhjk\right]_2} \ldots 
\int_{-\infty}^{\left[\qhjk\right]_p} 
log\left(1+\frac{\boldsymbol s^T{\bLambda_h}^{-1}\boldsymbol 
s}{\nu_h^{(k)}+d_h^{(k)}(\byj)}\right) \label{S1b}\\
	&	& \left[1+\frac{\boldsymbol s^T{\bLambda_h}^{-1}\boldsymbol 
	s}{\nu_h^{(k)}+d_h^{(k)}(\byj)}\right] 
	^{-\left(\frac{\nu_h^{(k)}}{2}+p\right)} d\boldsymbol u, \nonumber
\end{eqnarray} 
$\boldsymbol S_{2,hj}^{(k)}$ is a $p \times 1$ vector whose $r$th element is
\begin{eqnarray}
	&	&	\int_0^\infty \int_0^\infty \ldots \int_0^\infty u_r\; t_{p, 
\nu_h^{(k)}+p+2} \left(\boldsymbol u; \qhjk, 
\left(\frac{\nu_h^{(k)}+d_h^{(k)}(\byj)} 
{\nu_h^{(k)}+p+2}\right)\bDelta_h^{(k)} \right) d\boldsymbol u,
\label{S2}
\end{eqnarray}
 and $\boldsymbol S_{3,hj}^{(k)}$ is a $p \times p$ matrix whose $(r,s)$th 
 element is 
\begin{eqnarray} 
	&	&	\int_0^\infty \int_0^\infty \ldots \int_0^\infty u_r\; u_s\; t_{p, 
\nu_h^{(k)}+p+2} \left(\boldsymbol u; \qhjk, 
\left(\frac{\nu_h^{(k)}+d_h^{(k)}(\byj)} 
{\nu_h^{(k)}+p+2}\right)\bDelta_h^{(k)} \right) d\boldsymbol u, 
\label{S3}
\end{eqnarray}
where, for convenience of notation, $d\boldsymbol u$ is used to denote $du_1, 
du_2, \ldots, du_p$.

It is important to note that $S_{2,hj}^{(k)}$ and $S_{3,hj}^{(k)}$ are 
related to the first two moments of a truncated $p$-dimensional 
$t$-variate $\boldsymbol X_{hj}$. More specifically, let 
\begin{equation} 
\boldsymbol X_{hj} \sim tt_{p,\nu_h+p+2}\left(\qhjk, 
\left(\frac{\nu_h^{(k)}+\dhkyj} 
{\nu_h^{(k)}+p+2}\right)\bDelta_h^{(k)}, 
\mathbb{R}^+\right),
\nonumber
\end{equation} 
the truncated $t$-distribution as defined by (\ref{TTden}). Then 
\begin{align}
S_{2,hj}^{(k)} &= T_{p,\nu_h^{(k)}+p+2}\left(\qhjk; \boldsymbol 
0,\left(\frac{\nu_h^{(k)}+\dhkyj} {\nu_h^{(k)}+p+2}\bDelta_h^{(k)}\right)\right) 
E(\boldsymbol X_{hj}), \notag\\
\intertext{and}
S_{3,hj}^{(k)} &= T_{p,\nu_h^{(k)}+p+2}\left(\qhjk; \boldsymbol 0, 
\left(\frac{\nu_h^{(k)}+\dhkyj} {\nu_h^{(k)}+p+2}\bDelta_h^{(k)}\right)\right) 
E(\boldsymbol X_{hj}\boldsymbol X_{hj}^T), \notag
\end{align}
and hence (\ref{e3}) and (\ref{e4}) reduces to $\be_{3,hj}^{(k)} = e_{2,hj}^{(k)} E(\boldsymbol X_{hj})$ and $\be_{4,hj}^{(k)} = e_{2,hj}^{(k)} E(\boldsymbol X_{hj} \boldsymbol X_{hj}^T)$ respectively,
which can be implicitly expressed in terms of the parameters  $\qhjk$, $\dhkyj$, 
$\bDelta_h^{(k)}$, $\nu_h^{(k)}$ using results (\ref{EX}) and (\ref{EXX}). It is 
worth emphasizing that computation of $\be_{3hj}^{(k)}$ and $\be_{4hj}^{(k)}$ 
depends on algorithms for evaluating the multivariate $t$-distribution function, 
for which fast procedures are available. 
\newline  

In summary, the ECM algorithm is implemented as follows on the $(k+1)$th
iteration:

\textbf{E-step:} Given $\bPsi = \bPsi^{(k)}$, compute $\tau_{hj}^{(k)}$ using 
(\ref{TAU}), and $e_{1,hj}^{(k)}$, $e_{2,hj}^{(k)}$, $\be_{3,hj}^{(k)}$, and 
$\be_{4,hj}^{(k)}$ as described by (\ref{e1}), (\ref{e2}), (\ref{e3}), and 
(\ref{e4}) respectively, for $h=1, \ldots,g$ and $j=1,\ldots,n$. 
\newline

\textbf{M-step:} Update the estimate of $\bPsi$ by calculating for 
$h=1,\ldots, g$, the following estimates of the parameters in $\bPsi$,

\begin{align}
\bmu_h^{(k)} &= \frac{\sum_{j=1}^n 
	\tau_{hj}^{(k)} \left[e_{2,hj}^{(k)}\byj - 
	\bDelta_h^{(k)}\be_{3,hj}^{(k)}\right]} {\sum_{j=1}^n 
	\tau_{hj}^{(k)}e_{2,hj}^{(k)}}, \notag\displaybreak[0]\\
\bdelta^{(k+1)} &= \left(\bSigma_h^{(k)^{-1}} \odot 
	\sum_{j=1}^n \tau_{hj}^{(k)} \be_{4,hj}^{(k)}\right)^{-1}
	\mbox{DIAG}\left(\bSigma_h^{(k)^{-1}} 
	\sum_{j=1}^n \tau_{hj}^{(k)} (\by_j-\bmu_h^{(k)}) \be_{3,hj}^{(k)^T} \right),
	\notag\displaybreak[0]\\
	\intertext{and}
{\bSigma}_h^{(k+1)} &=	\frac{1} {\sum_{j=1}^n 
		\tau_{hj}^{(k)}} \sum_{j=1}^n \tau_{hj}^{(k)} 
		\left[\bDelta_h^{(k+1)}\be_{4,hj}^{(k)^T}\bDelta_h^{(k+1)^T} \right. 
		 \left(\byj-\bmu_h^{(k+1)}\right) \be_{3,hj}^{(k)^T} 
	\bDelta_h^{(k+1)} \notag\\
	&	- \bDelta_h^{(k+1)}\be_{3,hj}^{(k)} 
	\left(\byj-\bmu_h^{(k+1)}\right)^T 
	\left. + \left(\byj-\bmu_h^{(k+1)}\right) \left(\byj-\bmu_h^{(k+1)}\right)^T 
	e_{2,hj}^{(k)}\right]. \notag
\end{align}

	An update $\nu_h^{(k+1)}$ of the degrees of freedom is obtained by solving 
	iteratively the equation
	\begin{equation}
	\log\left(\frac{\nu_h^{(k+1)}}{2}\right) - 
	\psi\left(\frac{\nu_h^{(k+1)}}{2}\right) = \frac{\sum_{j=1}^n 
	\left[\tau_{hj}^{(k)}\left(e_{2,hj}^{(k)}-e_{1,hj}^{(k)}-1\right)\right]} 
	{\sum_{j=1}^n \tau_{hj}^{(k)}}.
	\nonumber
	\end{equation} 

A program for implementing this EM algorithm has been written in R.

\section{The Empirical Information Matrix}
\label{sec:6}

We consider an approximation to the asymptotic covariance matrix of the ML 
estimates using the inverse of the empirical information matrix (Basford et al., 
1997). The empirical information matrix is given by \begin{equation}
I_e\left(\bPsi; \boldsymbol y\right) = \sum_{j=1}^n \boldsymbol s\left(\byj; 
\hat{\bPsi}\right) \boldsymbol s^T\left(\byj; \hat{\bPsi}\right),
\label{Ie}
\end{equation}
where $s\left(\byj; \hat{\bPsi}\right) = E_{\hat{\bPsi}} \left\{\partial \log 
L_{cj}\left(\bPsi\right)/\partial\bPsi \mid\byj\right\}$  $(j=1,\ldots,n)$ are the 
individual scores, consisting of
\begin{align} 
&(s_{j,\pi_1}, \ldots, s_{j,\pi_{g-1}}, s_{j,\bmu_1}, 
\ldots, s_{j,\bmu_g}, s_{j, \bdelta_1}\notag\\ 
&\ldots, s_{j,\bdelta_g}, s_{j,\bSigma_1}, \ldots, s_{j,\bSigma_g}, s_{j,\nu_1}, 
\ldots, s_{j,\nu_g} ). \nonumber
\end{align}
We let $L_{cj}\left(\bPsi\right)$ denote the complete-data log 
likelihood formed from the single observation $\byj$. An estimate of 
the covariance matrix of $\hat{\bPsi}$ is given by taking the inverse 
of (\ref{Ie}). After some algebraic manipulations, one can show that the elements 
of $\boldsymbol s\left(\byj; \hat{\bPsi}\right)$ are given 
by the following explicit expressions:

\begin{align}
s_{j,\pi_h}	&=	\frac{\tau_{hj}}{\pi_h} - \frac{\tau_{gj}}{\pi_g},  
\notag\displaybreak[0]\\
s_{j,\bmu_h}	&=	\tau_{hj}\hat{\bSigma}_h^{-1} \left[e_{2,ij}\left(\byj- 
\hat{\bmu}_h\right) - \hat{\bDelta}_h \be_{3,ij}\right], 
\notag\displaybreak[0]\\
s_{j,\bSigma_h}	&=	\textstyle \frac{1}{2} \tau_{hj} 
\left[\left(\byj-\hat{\bmu}_h\right)\left(\byj-\hat{\bmu}_h\right)^T - 
\left(\byj-\hat{\bmu}_h\right)\be_{3,hj}^T\hat{\bDelta}_h 
\right. \notag\\
	&	\left. - \hat{\bDelta}_h\be_{3,hj}\left(\byj-\hat{\bmu}_h\right) + 
	\hat{\bDelta}_h\be_{3,ij}\left(\byj-\hat{\bmu}_h\right) + 
	\hat{\bdelta}_h\be_{4,hj}^T\hat{\bDelta}_h\right] \hat{\bSigma}_h^{-1} 
	\notag\\
	&	- \textstyle\frac{1}{2}\tau_{hj} \hat{\bSigma}_h^{-1}, 
	\notag\displaybreak[0]\\
s_{j,\bdelta_h} &=	\tau_{hj} \left[ 
	\mbox{diag}\left(\hat{\bSigma}_h^{-1} \left(\byj-\hat{\bmu}_h\right)
	\right) \be_{3,hj} - \left(\hat{\bSigma}_h^{-1} \odot \be_{4,hj}\right)
	\hat{\bdelta}_h \right],	\notag\displaybreak[0]\\
s_{j,\nu_h} &=	\textstyle\frac{1}{2} \tau_{hj} 
\left[\log\left(\textstyle\frac{1}{2} \hat{\nu}_h\right) + 1 + e_{1,hj} - 
\psi\left(\textstyle\frac{1}{2}\hat{\nu}_h\right) - e_{2,hj}\right]. \notag
\end{align}

\section{Examples}
\label{sec:7}

In this section, we fit the FM-MST model to three real data sets to demonstrate 
its usefulness in analyzing and clustering multivariate skewed data. In the first 
example, we focus on the flexibility of the FM-MST model in capturing the 
asymmetric shape of flow cytometric data. The next example illustrates the 
clustering capability of the model. In the final example, we demonstrate the 
computational efficiency of the proposed algorithm.

\subsection{Lymphoma Data}
\begin{figure}[tbp]
	\centering
 	\includegraphics[width=1.00\textwidth]{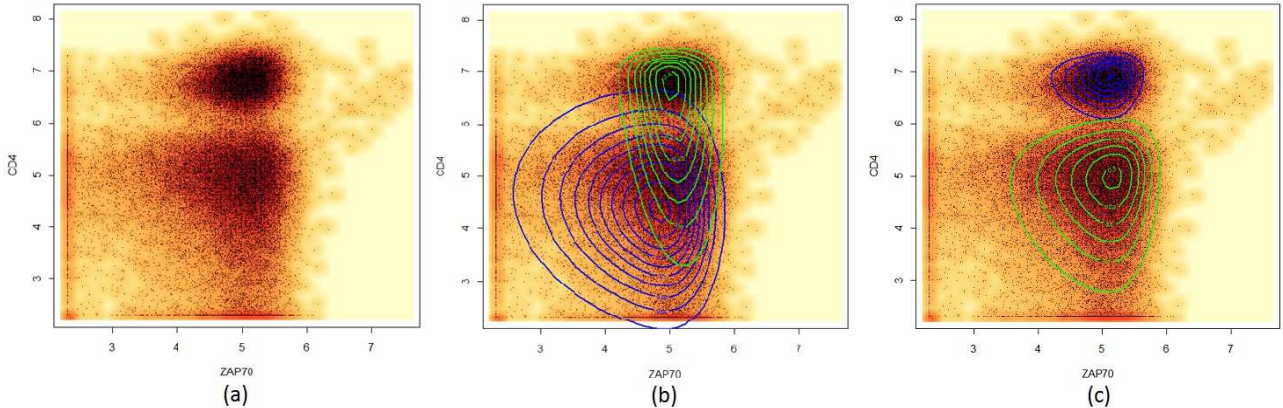}
	\caption{\emph{Mixture modelling of a reduced subset of prephosphorylation 
	T cell population. Bivariate skew $t$-mixtures were fitted to the data 
	restricted in two dimensions CD45 and ZAP70. (a) Hue intensity plot of the 
	Lymphoma data set; (b) the contours of the component densities in the fitted 
	two-component skew $t$-mixture model FM-SNI-ST using the R package 
	\texttt{mixsmsn}; (c) the fitted component contours of the two-component FM-MST 
	model.}}
	\label{fig1}
\end{figure}

We consider a subset of the T-cell phosphorylation data collected by 
Maier et al. (2007). In the original data, blood samples from 30 subjects were 
stained with four fluorophore-labeled antibodies against CD4, CD45RA, 
SLP76(pY128), and ZAP70(pY292) before and after an anti-CD3 stimulation. In this 
example, we focus on a reduced subset of the data in two variables -- CD4 and 
ZAP70. This bivariate sample (Figure \ref{fig1}) is apparently bimodal and 
exhibits asymmetric pattern. Hence we fit a two-component FM-MST model to the 
data. More specifically, the fitted model can be written as
\begin{equation}
f_2\left(\byj; \bPsi\right) = \pi_1 f_2\left(\byj;\bmu_1, \bSigma_1, \bdelta_1, 
\nu_1\right) + \left(1-\pi_1\right) f_2\left(\byj; \bmu_2, \bSigma_2, \bdelta_2, 
\nu_2\right),
\nonumber
\end{equation}
where 
\begin{equation}
\bmu_i = \left(\mu_{i,1}, \mu_{i, 2}\right)^T, \;
\bSigma_i = \left(\begin{array}{cc} 
\sigma_{i,11} & \sigma_{i,12} \\ \sigma_{i,12} & \sigma_{i,22}  
\end{array}\right), 
\bdelta_i = \left(\delta_{i,1}, \delta_{i,2}\right)^T 
(i = 1, 2). \nonumber
\end{equation}

For comparison, we include the fitting of a two-component mixture of skew 
$t$-distributions from the skew-normal independent (SNI) family (Lachos, Ghosh, 
and Arellano-Valle, (2010)), hereafter named the FM-SNI-ST model. The 
estimated FM-SNI-ST density can be computed from the R package \texttt{mixsmsn} 
(Cabral, Lachos, and Prates, (2012)). Note that the MST distribution is 
different to the SNI-ST distribution since the skewing function is not of 
dimension one. Note also that the SNI-ST distribution is equivalent to the restricted MST distribution (\ref{MST_Sam}) after  reparametrization.
Moreover, under the FM-SNI-ST settings, the correlation structure 
of $\boldsymbol Y$ will also be dependent on the skewness parameter, whereas for 
the FM-MST distributions the covariance structure is not affected by $\bdelta$. 
The contours of the fitted SNI-ST and MST component densities are 
depicted in Fig~\ref{fig1}(b) and Fig~\ref{fig1}(c), respectively. To better 
visualize the shape of the fitted models, we display the estimated densities of 
each component instead of the mixture contours. It can be seen that the FM-MST 
model provides a noticeably better fit. From a clustering point of view, the 
FM-MST model also shows better performance as it is able to separate the two 
clusters correctly. Moreover, it adapts to the asymmetric shape of each cluster 
more adequately. Thus the superiority of FM-MST model is evident in dealing with 
asymmetric and heavily tailed data in this data set.

\subsection{GvHD Data}

\begin{figure}[tbp]
	\centering
 	\includegraphics[width=1.00\textwidth]{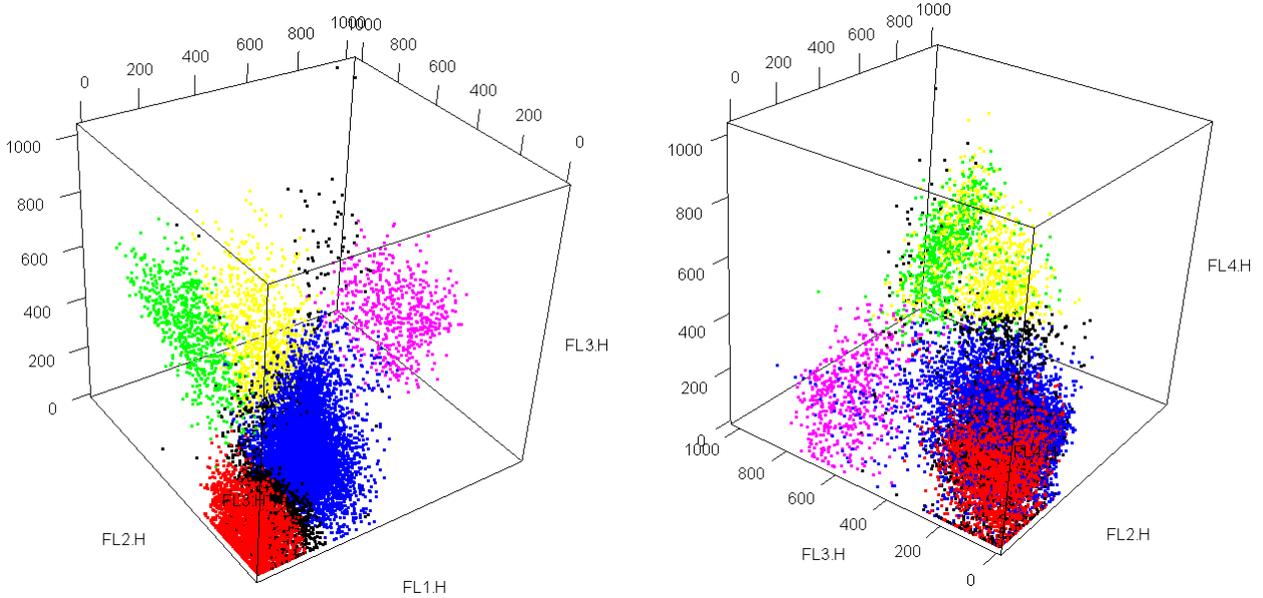}
	\caption{GvHD data set: \emph{Expert manual clustering of a population of 
	13773 cells stained with four fluorescence reagents -- CD4 FITC (FL1-H), 
	CD8$\beta$ PE (FL2.H), CD3 PerCP (FL3.H) and CD8 APC (FL4.H).}}
	\label{fig2}
\end{figure}

Our second example concerns a data set collected by Brinkman et al. (2007), where 
peripheral blood samples were collected weekly from patients following blood and 
bone marrow transplant. The original goal was to identify cellular signatures that 
can predict or assist in early detection of Graft versus Host Disease (GvHD), a 
common post-transplantation complication in which the recipient's bone marrow was 
attacked by the new donor material. Samples were stained with four fluorescence 
reagents: CD4 FITC, CD8$\beta$ PE, CD3 PerCP, and CD8 APC. Hence we fit a 
4-variate FM-MST model to a case sample with a population of 13773 cells. The 
data set is shown in Figure \ref{fig2}, where cells are displayed in five 
different colours according to a manual expert clustering into five clusters. In 
addition, we include the results for the FM-SNI-ST model and the restricted MST mixture model introduced in Section \ref{sec:2.1} (equation \ref{MST_Sam}), 
hereafter denoted by FM-RMST. 

We compare the performance of the three models FM-MST, FM-SNI-ST, and FM-RMST in 
assigning cells to the expert clusters. Manual gating suggests there are five 
clusters in this case sample. Hence we applied the algorithm for the fitting of 
each model with $g$ predefined as $5$. For a fair comparison, we started the three 
algorithms using the same initial values. The initial clustering is based on 
$k$-means.The degrees of freedom are set to be identical for all components 
for the first iteration and assigned a relatively large value. A similar strategy 
was described in Lin (2010).  

To assess the performance of these three algorithms, we take the manual expert 
clustering as being the `true' class membership and we calculated the error rate 
of classification against this benchmark result with dead cells removed, measured 
by choosing among the possible permutations of class labels the one that gives the 
highest value. 

As anticipated, the optimal clustering result was given by the FM-MST model. It 
achieved the lowest misclassification rate. The FM-SNI-ST model has a higher 
number of misallocations. The FM-RMST model has a disappointing performance in 
terms of clustering. Its error rate is almost double that of its competitors. It 
is worth pointing out that both the FM-MST and FM-RMST models have 99 free 
parameters, while the FM-SNI-ST model has 95 parameters. It is evident that these 
two restricted models have inferior performance. This reveals some evidence of the 
extra flexibility offered by the more general FM-MST model. 

\begin{table*}
	\centering
	\caption{Clustering error rates for various multivariate skew $t$ mixture 
	models on the GvHD data set.}
		\begin{tabular}{ccc}
			\hline
			Model	&	Error rate	&	Number of free parameters\\	
			\hline
			FM-MST	& \bf{0.0875} &	99 \\
			FM-SNI-ST	&	0.13078 & 95  \\
			FM-RMST	&	0.20700 & 99 	\\	
			\hline
		\end{tabular}
	\label{tab1}
\end{table*}

\subsection{AIS Data} 

We now illustrate the computational efficiency of our exact implementation of the 
E-step of the EM algorithm as in Section \ref{sec:5}. 
We denote this version of the EM algorithm 
with the exact E-step as EM-exact. In addition, we consider the EM alternative 
with a Monte Carlo (MC) E-step as given by Lin (2010), which is denoted by 
EM-MC. Since both models are based on the same characterization of the 
multivariate skew $t$-distribution defined by \Sahu, it is appropriate to compare 
their computation time. We assess their time performance on the well-analyzed 
Australian Institute of Sport (AIS) data, which consists of $p=13$ measurements 
made on $n=202$ athletes. As in Lin (2010), we limit this illustration to a 
bivariate subset of two variables -- body mass index (BMI) and the percentage of 
body fat (Bfat). As noted by Lin (2010), these data are apparently bimodal. 
Hence a two-component mixture model is fitted to the data set. 

A summary of the results are listed in Table \ref{tab2}. Also reported there are 
the values of the log-likelihood, the Akaike information criterion (AIC) 
(Akaike, 1974) and the Bayesian information criterion (BIC) (Schwarz, 1978) 
defined by
\begin{equation}
\mbox{AIC} = 2m - 2L\left(\bPsi\right) \; \mbox{and} \; 
\mbox{BIC} = m \log n - 2 L\left(\bPsi\right),
\end{equation}
respectively, where $L\left(\bPsi\right)$ is the value of the log likelihood at 
$\bPsi$, $m$ is the number of free parameters, and $n$ is the sample size. Models 
with smaller AIC and BIC values are preferred when comparing different fitted 
results. The best value from each criterion are highlighted in bold font in Table 
\ref{tab2}. For this illustration, the EM-MC E-step is undertaken with $50$ 
random draws as recommended by Lin (2010). 
Note that the degrees of freedom is not restricted to be the same for the two components. 
The gender of each individual in this 
data set is also recorded, thus enabling us to evaluate the error rate of binary 
classification for the two methods. 

Not surprisingly, the model selection criteria favour the EM-exact algorithm. Not 
only did it achieve lower AIC and BIC values, the computation time is 
remarkably lower than its competitor. It is more than five times faster than the 
EM-MC alternative.  

\begin{table*}
	\centering
	\caption{Computation time and clustering error rates for two different 
implementations of the EM algorithm for the multivariate skew $t$ mixture models on
the AIS data set. For EM-exact, the E-step is implemented exactly as described in 
Section \ref{sec:5}. As an alternative, the EM algorithm was implemented with a 
Monte Carlo E-step, EM-MC, as in Lin (2010). Time is measured in seconds.}
	\begin{tabular}{cccccc}
			\hline	
			Model	&	\multicolumn{2}{c}{EM-exact}	&&	
			\multicolumn{2}{c}{EM-MC}	\\\cline{2-3}\cline{5-6}
			Component	& 1	&	2	&	&	1	&	2	\\		
			\hline
			$\pi$	&	0.44	&	0.56	&&	0.59	&	0.41	\\
			$\mu_{i1}$	&	19.74	&	21.83 &&	19.89	&	22.47	\\
			$\mu_{i2}$ 	&	15.99	&	5.89	&&	15.50	&	7.30	\\	
			$\Sigma_{i,11}$	&	3.03	&	3.16	&&	2.96	&	3.23	\\
			$\Sigma_{i,12}$	&	7.71	&	0.54	&&	6.17	&	1.34	\\
			$\Sigma_{i,22}$	&	2.36	&	0.11	&&	25.80	&	2.14	\\
			$\delta_{i1}$ 	&	3.34	&	1.44	&&	2.72	&	0.71	\\
			$\delta_{i2}$	&	3.15	&	3.76	&&	2.22	&	1.13	\\
			$\nu$ &	42.05	&	3.82	&&	23.98	&	25.93	\\
			$L\left(\bPsi\right)$ 
			&	\multicolumn{2}{c}{-1077.257}	&& \multicolumn{2}{c}{-1088.066}	\\
			AIC	&	\multicolumn{2}{c}{\bf{2188.514}}	&&	\multicolumn{2}{c}{2207.956}	\\
			BIC	&	\multicolumn{2}{c}{\bf{2244.755}}	&&	\multicolumn{2}{c}{2264.197}	\\
			error rate	&	\multicolumn{2}{c}{0.0792}	&&	
			\multicolumn{2}{c}{0.0891}	\\
			time	&	\multicolumn{2}{c}{\bf{64.63}}	&&	\multicolumn{2}{c}{349.9}	\\	
			\hline
		\end{tabular}
	\label{tab2}
\end{table*}

\section{Computation Time and Accuracy for E-step}
We now proceed to two interesting experiments for evaluating the computational 
cost and accuracy of using the EM-exact and EM-MC algorithms on high-dimensional 
data. As pointed out previously, the main computational cost for EM-exact is 
evaluating the multivariate $t$-distribution function. Calculation of the first 
two moments of a $p$-variate truncated $t$-distribution requires the evaluation of 
two  $T_p(\cdot)$ functions, $p$ evaluations of $T_{p-1}(\cdot)$, and 
$\textstyle\frac{1}{2} p (p-1)$ evaluations of $T_{p-2}(\cdot)$. Hence, the 
computation time will increase substantially with the number of dimensions. 
However, with the EM-exact algorithm, accuracy can be compromised for time. 

\begin{figure}[tbp]
	\centering
 	\includegraphics[width=1.0\textwidth]{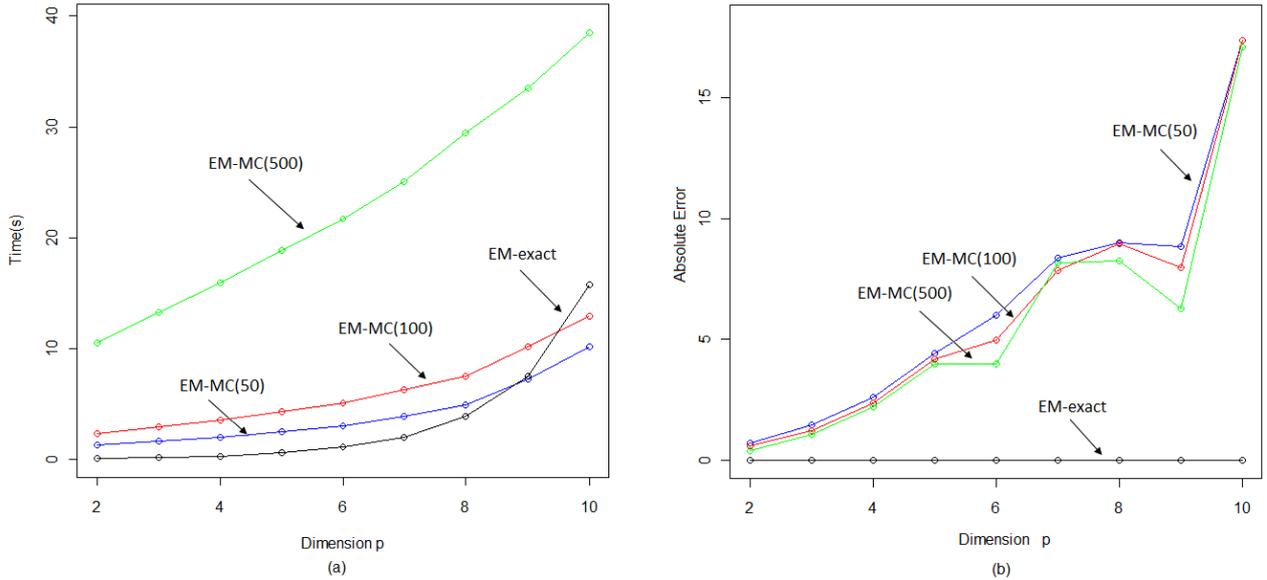}
	\caption{\emph{Comparison of performance of the EM-MC and EM-exact methods on 
	a subset of 100 samples from the Brain Tumor data. Green line: EM-MC with 500 
	draws, red line: EM-MC with 100 draws, blue line: EM-MC with 50 draws, black 
	line: EM-exact. (a) Typical computation time for E-step on a sample of 100 data 
	in various dimensions. (b) Total absolute error of E-step for one data point.}}
	\label{fig3}
\end{figure}
  
We sampled 100 data from a Brain Tumor dataset supplied by Geoff Osborne from the 
Queensland Brain Institute at the University of Queensland. In both experiments we 
varied the dimension $p$ of the sample. The graph in Figure \ref{fig3}(a) 
shows the typical CPU time per each E-step iteration for various dimensions $p$ of 
the data; EM-MC$(m)$ represents the EM-MC algorithm with $m$ random draws using 
the Gibbs sampling approach described in Lin (2010). It is worth noting that in 
both experiments EM-exact is evaluated with a default tolerance of at least 
$10^{-6}$. As seen in Figure \ref{fig3}, EM-exact is the fastest among the four 
versions of the E-step for low dimensions. For example, at $p=2$, EM-exact at 
least 25 times faster than EM-MC(50). It is important to note that although 
EM-MC(50) is slightly faster than EM-exact at higher dimensions, EM-exact produces 
results to a significantly higher accuracy, while EM-MC requires a large number of 
draws to achieve comparable results. We note that in our simulations, for example, 
at $p=7$, 50 draws is insufficient to achieve acceptable estimates. Preliminary 
results suggests that at least 500 draws is required to generate reasonable 
approximations when $p$ is greater than 6. In this case, EM-exact is at least ten 
times quicker. Furthermore, EM-exact also has an additional advantage over the 
EM-MC alternative in that its results are reproducible.   

To compare the accuracy of the EM-exact and EM-MC algorithms, we compute the total 
absolute error against the baseline EM-exact with a maximum tolerance of 
$10^{-18}$. For each of the EM-MC$(m)$ algorithms, the average total absolute 
error of 100 trials is used. For EM-exact, the default tolerance is set to 
$10^{-6}$. The results are shown in Figure \ref{fig3}(b). Not surprisingly, the 
absolute error of the EM-MC algorithm is significantly higher than that of the 
EM-exact algorithm. It can be observed that the absolute error is very high even 
for EM-MC(500). At $p=10$, for example, EM-exact is at least 30000 times more 
accurate and takes less than half the time required for EM-MC(500).  
 
It is important to emphasize that as the dimension $p$ of the data increases, 
EM-MC requires considerably more draws to provide a comparable (and acceptable) 
level of accuracy as EM-exact, which can be computationally intensive. Hence we 
advocate the use of EM-exact, especially for applications involving high 
dimensional data.  

\section{Concluding Remarks}

We have described an exact EM algorithm for evaluating the parameters of a general 
multivariate skew $t$-mixture model. This model has a more general 
characterization than various alternative versions of the skew $t$-distribution 
available in the literature and hence offers greater flexibility in capturing the 
asymmetric shape of skewed data. 

Our proposed method is based on reduced analytical expressions for the E-step 
conditional expectations, which can be formulated in terms of the first and second 
moments of a multivariate truncated $t$-distribution. The latter can then be 
expressed further in terms of the distribution function of the multivariate 
central $t$-distribution for which fast algorithms capable of producing highly 
accurate results already exist. It is demonstrated to have a marked advantage over 
the EM algorithm with a Monte Carlo E-step. To achieve comparable accuracy to that 
of the EM algorithm with the E-step implemented using the above numerical 
approach, the version of the algorithm with a Monte Carlo E-step would require a 
large number of draws, which would be computationally expensive.

\section*{Acknowledgments}
This work is supported by a grant from the Australian Research Council. Also, we 
would like to thank Professor Seung-Gu Kim for comments and corrections, and Dr Kui (Sam) Wang for his helpful discussions on this topic.

\end{document}